\documentclass[mathleft
]{an}
\usepackage{graphicx}
\usepackage{times}
\begin{document}

\Pagespan{789}{}
\Yearpublication{2006}%
\Yearsubmission{2005}%
\Month{11}%
\Volume{999}%
\Issue{88}%

\title{PAOLO: a Polarimeter Add-On for the LRS Optics at a Nasmyth focus of the TNG\thanks{Based on observations made with the Italian Telescopio Nazionale Galileo (TNG) operated on the island of La Palma by the Fundaci—n Galileo Galilei of the INAF (Istituto Nazionale di Astrofisica) at the Spanish Observatorio del Roque de los Muchachos of the Instituto de Astrofisica de Canarias"}}

\author{S. Covino\inst{1}\fnmsep\thanks{Corresponding author:
  \email{stefano.covino@brera.inaf.it}\newline}
\and  E. Molinari\inst{2,3}
\and P. Bruno\inst{4}
\and M. Cecconi\inst{2}
\and P. Conconi\inst{1}
\and P. D'Avanzo\inst{1}
\and L. di Fabrizio\inst{2}
\and D. Fugazza\inst{1}
\and M. Giarrusso\inst{5,4}
\and E. Giro\inst{6}
\and F. Leone\inst{5,4}
\and V. Lorenzi\inst{2}
\and S. Scuderi\inst{4}
}
\titlerunning{The PAOLO polarimeter}
\authorrunning{S. Covino et al.}
\institute{
INAF / Brera Astronomical Observatory, Via Bianchi 46, 23807, Merate (LC), Italy
\and 
INAF / Fund. Galileo Galilei, Rambla Jos\'e Ana Fern\'andez Perez 7, 38712 Bre\~{n}a Baja (La Palma), Canary Islands, Spain
\and
INAF / IASF--MI, Via E. Bassini 15, 20133, Milano, Italy
\and
INAF / Osservatorio Astrofisico di Catania, Via S. Sofia 78, 95123, Catania, Italy
\and
Universit\`a di Catania, Dipartimento di Fisica e Astronomia, Sezione Astrofisica, Via S. Sofia 78, 95123, Catania, Italy
\and
INAF / Padova Astronomical Observatory, Vicolo dell'Osservatorio 5, 35122, Padova, Italy
}

\keywords{instrumentation: polarimeters -- methods: data analysis -- techniques: polarimetric}

\abstract{%
We describe a new polarimetric facility available at the Istituto Nazionale di AstroFisica / Telescopio Nazionale Galileo  at La Palma, Canary islands. This facility, PAOLO (Polarimetric Add-On for the LRS Optics), is located at a Nasmyth focus of  an alt-az telescope and requires a specific modeling in order to remove the time- and pointing position-dependent instrumental polarization. We also describe the opto-mechanical structure of the instrument and its calibration and present early examples of applications.
}

\maketitle

\section{Introduction}
Polarimetry is a powerful diagnostic tool for studying astrophysical sources. Radiation mechanisms that produce similar radiation output can be disentangled by means of their polarization signatures. Also, polarization provides unique insights into the geometry of unresolved sources, hidden in the integrated light, even at cosmological distances.

Polarization and wavelength are the bits of information attached to every photon that reveal the most about its creation and subsequent history. In an observational science such as astronomy, polarimetry is especially important because it goes directly to the heart of the problem, i.e. the underlying physical process. The elementary processes include, among others:

\begin{itemize}
\item Scattering (e.g. by dust particles in the interstellar medium).
\item Absorption by aligned, intrinsically asymmetric particles (e.g. in molecular clouds).
\item Scattering by asymmetrically distributed particles (e.g. in disks).
\item Coherent scattering in spectral lines (e.g .in magnetospheres).
\item Emission or scattering by asymmetric sources (e.g .supernova explosions).
\item Cosmic magnetic and electric fields through observation of Zeeman-, Stark- and Paschen-Pack- effects (e.g. on stellar surfaces).
\item Cyclotron or synchrotron radiation, inverse-Compton process and relativistic jets (e.g. Active Galactic Nuclei, AGNs; Gamma-Ray Bursts, GRBs; etc.).
\item Reflection (e.g. by extra-solar planets).
\item Quantum gravity modification of standard dispersion relation (e.g. for cosmological sources).
\end{itemize}

The increasing importance of polarimetry in many astrophysical contexts is witnessed by the proportional increase in recent years 
of the number of scientific publications based on optical/NIR polarimetric observations at the European Southern Observatory (ESO)\footnote{http://www.eso.org} 
(Schmid 2008). 

With these considerations in mind, we designed, built, commissioned and calibrated a polarimetric add-on for the DOLORES\footnote{http://www.tng.iac.es/instruments/lrs/} (Device Optimized for the LOw RESolution)
instrument, LRS in short) at the INAF / Telescopio Nazionale Galileo (TNG)\footnote{http://www.tng.iac.es}. The main scientific goal of our project is to deliver a flexible all-purpose polarimetric facility hosted by a well calibrated and reliable instrument with a particular attention to provide polarization measurements of rapidly variable sources as GRB afterglows, blazars, and in general any high-energy transients.

In this paper we describe the technical design of the instrument in Sect.\,\ref{sec:des} and the polarimetric model adopted for PAOLO calibration in Sect\,\ref{sec:mod}.
Examples of typical PAOLO observing runs are briefly proposed in Sect.\ref{sec:exa}.

\section{Opto-mechanical design}
\label{sec:des}

\begin{figure}
\includegraphics[width=\columnwidth]{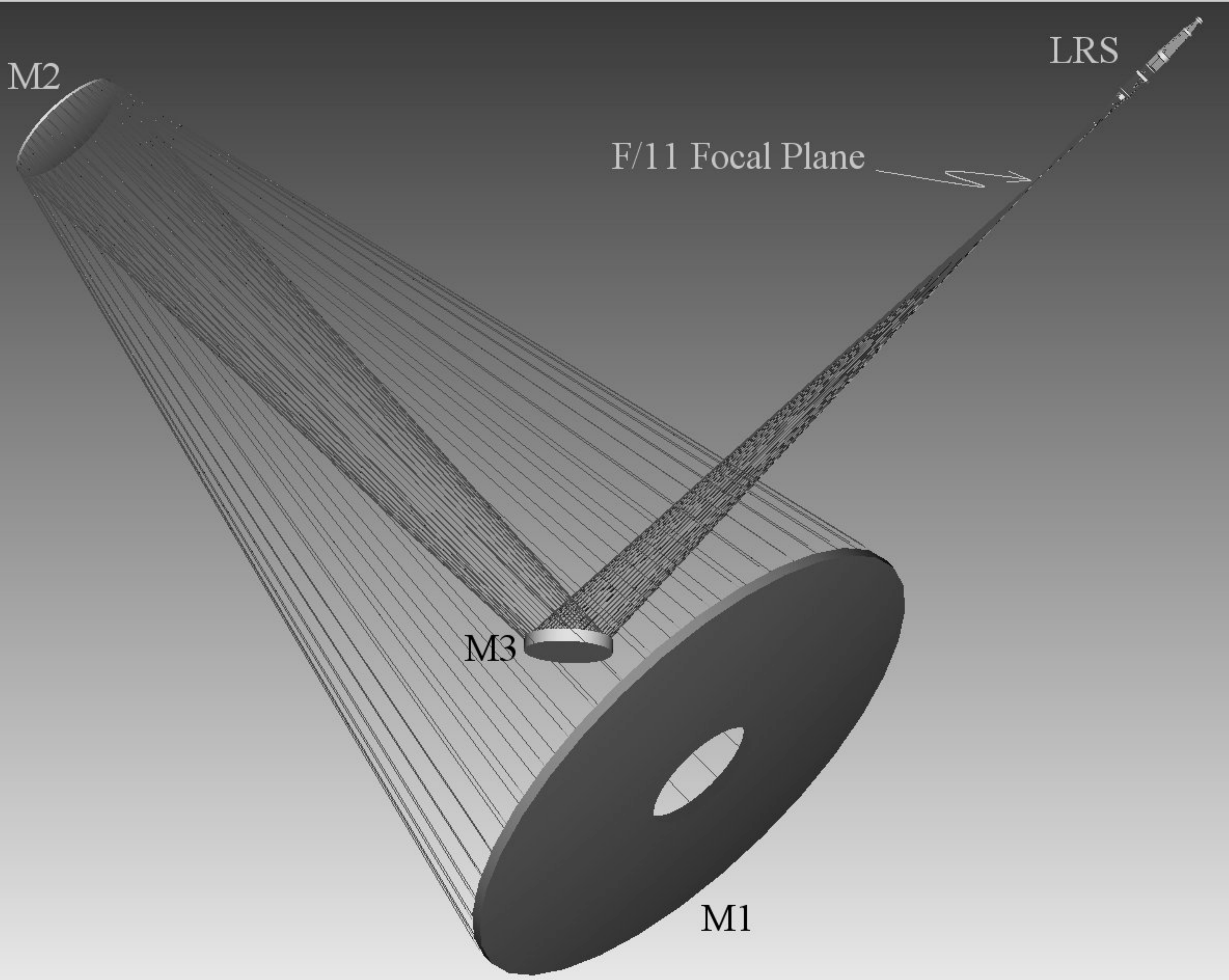}
\caption{Primary (M1) and secondary (M2) mirrors of the TNG. The M3 flat mirror can rotate by 180$^\circ$ to direct light to scientific instruments sited in Nasmyth A or Nasmyth B room.}
\label{fig:op}
\end{figure}

PAOLO is integrated in the LRS instrument, which is mounted on the Nasmyth B interface of the TNG. Primary (M1) and secondary (M2) mirrors of the TNG form an optical system with a 3580\,mm aperture and a 38500\,mm equivalent focal length with a corrected $15 \time15$\,arcmin$^2$ Field of View (FoV). The corresponding F/11 Focal Plane (FP) matches the LRS one (see Fig.\,\ref{fig:op}).

\begin{figure}
\includegraphics[width=\columnwidth]{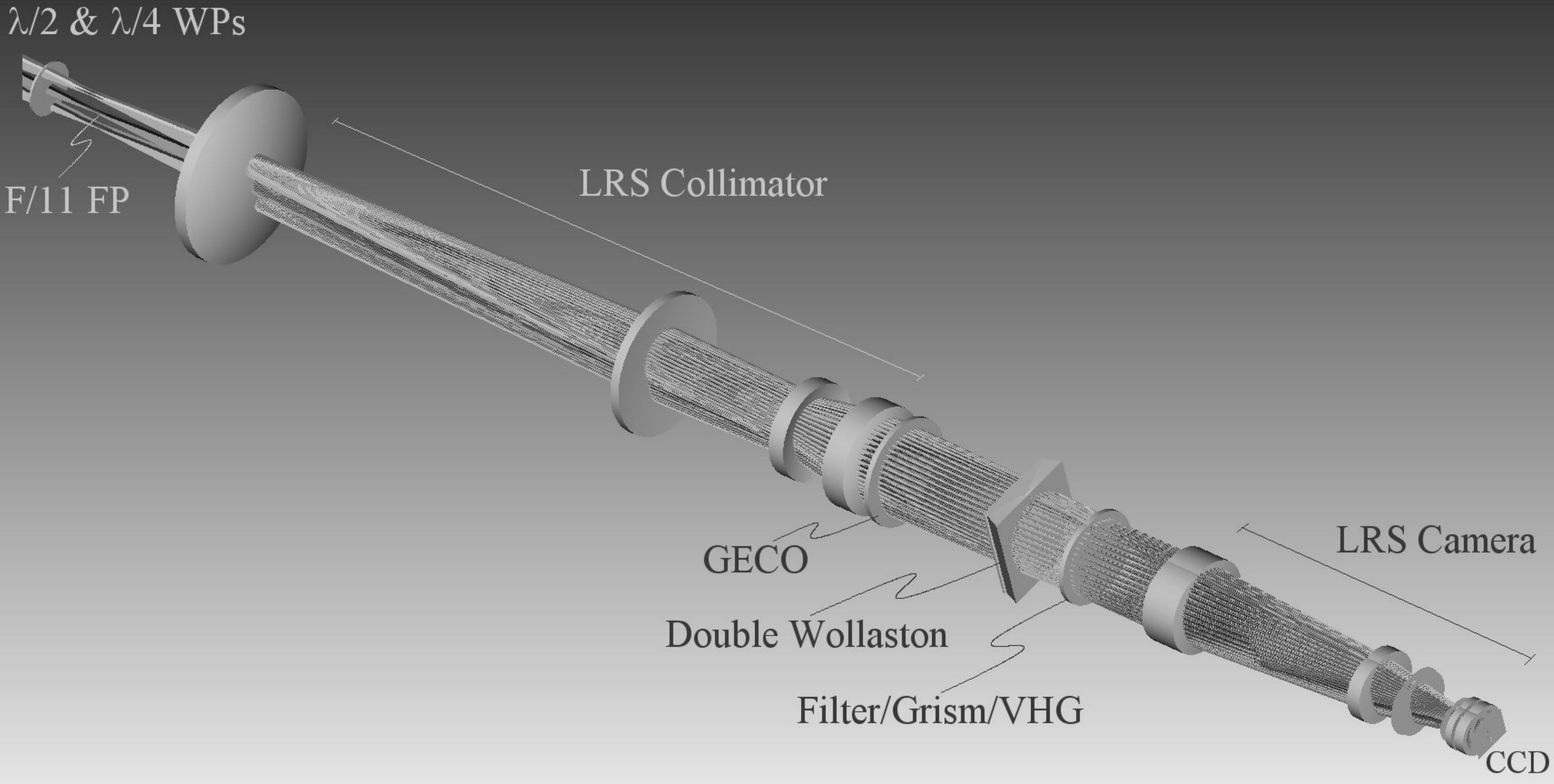}
\caption{PAOLO integrated in LRS setup at TNG.}
\label{fig:in}
\end{figure}

A set of six custom lenses set collimates the beam coming from the F/11 FP (see Fig.\,\ref{fig:in}). A Gravitational Eccentric Correction Optics (GECO; Conconi et al. 2002) then reduces the beam tilting caused by mechanical flexures during FoV derotation by the mechanical derotator. The collimated beam can then either a) pass through a filter mounted on a filter wheel or b) pass through a grism/volume holographic grating mounted on a grism wheel. A set of seven custom lenses set following the grism wheel forms the LRS camera which focuses the FoV or the spectra onto the $2048 \times 2048\ {\rm E2V} 4240$ thinned back-illuminated, deep-depleted, Astro-BB coated CCD with a pixel size of 13.5\,$\mu$m (0.252\,arcsec/pixel plate scale).

To optimize the trade-off between scientific goals and technical feasibility, we decided to mount the polarizer (a Double Wollaston, DW) on the filter wheel and the filters (for photo-polarimetry) and the grism/volume holographic gratings on the grism wheel. A rotating $\lambda$/2 or, alternatively, $\lambda$/4 waveplate can be set before the F/11 FP in the optical path to allow circular polarimetry measurement.

\begin{figure}
\centering
\begin{tabular}{cc}
\includegraphics[height=7cm]{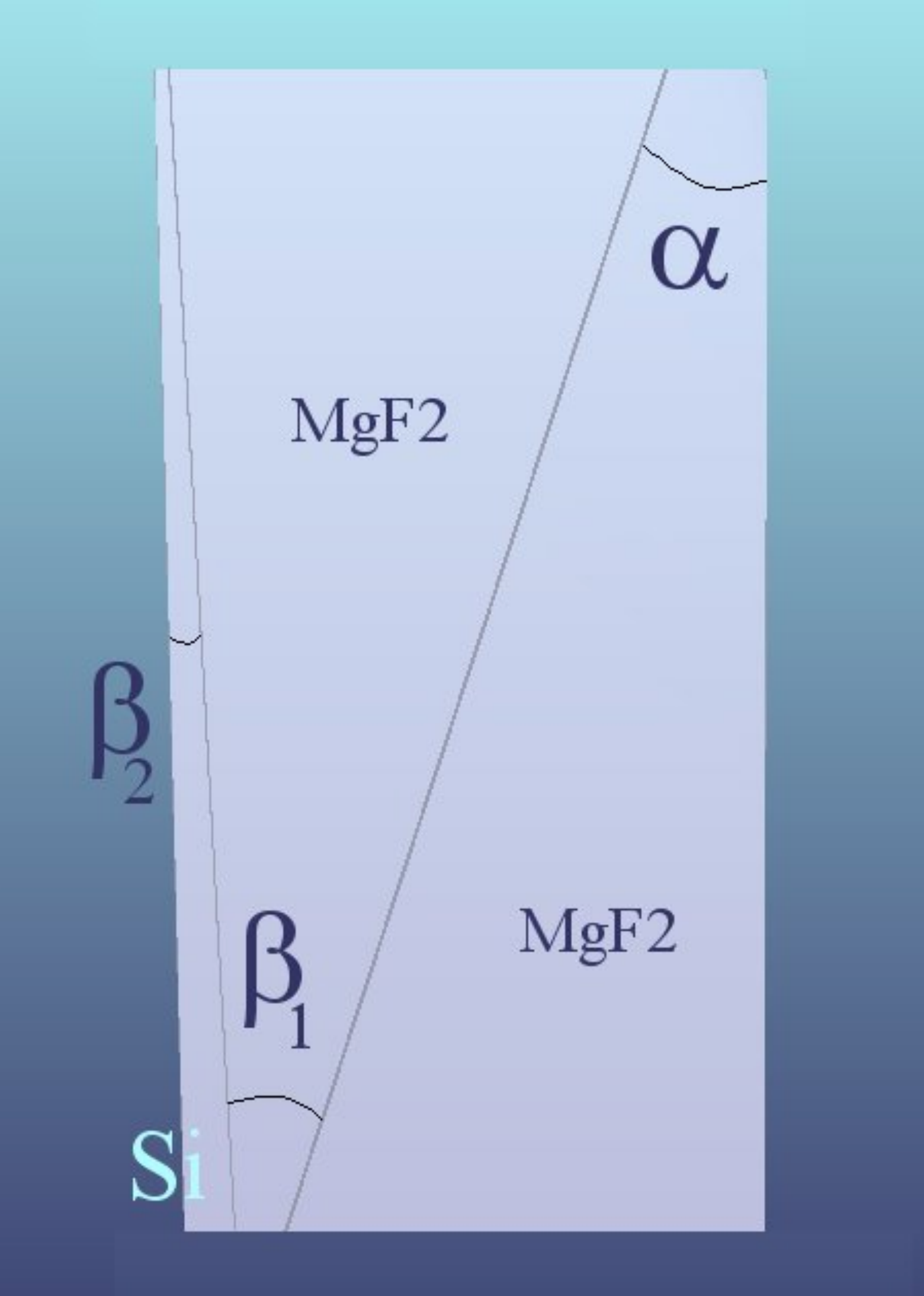} & 
\includegraphics[height=7cm]{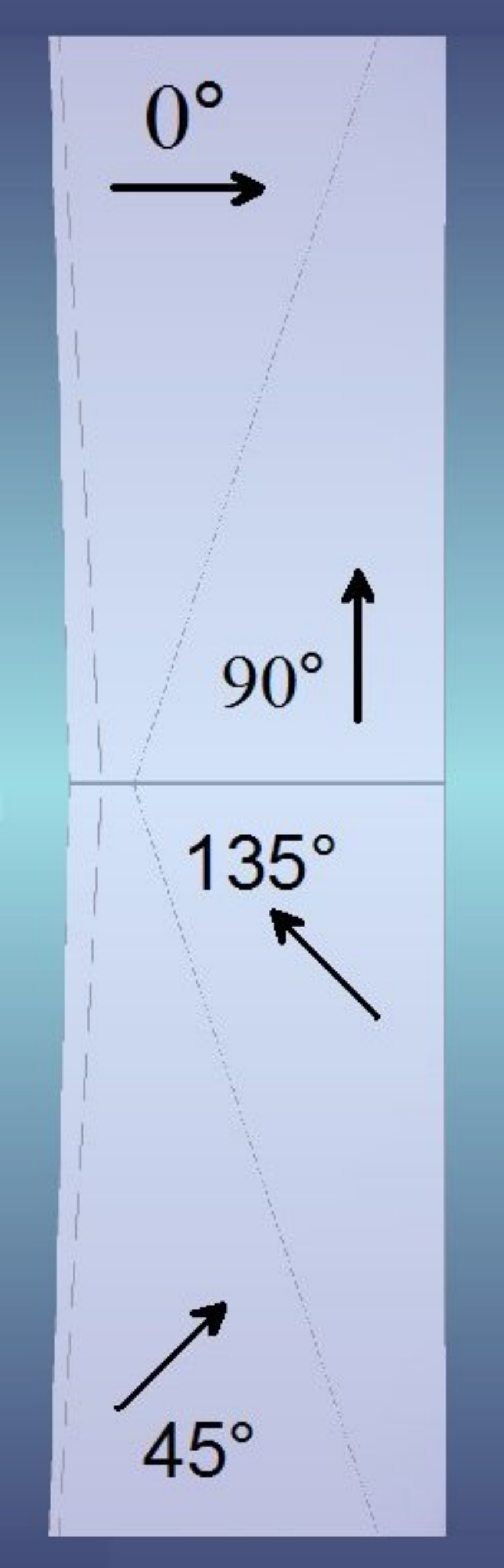} 
\end{tabular}
\caption{One of the two Wollaston (left side) forming the Double Wollaston polarizer (right side) glued along the center line. The front Si wedge with a $\beta_1$ angle allows for tilting the collimated beam to separate the FoV images corresponding to the two polarization state pairs (a pair for each Wollaston). $\alpha =18.00^\circ$, $\beta_1 = 3.41^\circ$ and $\beta_2 = 1.80^\circ$.}
\label{fig:DW}
\end{figure}

The DW opto-mechanical design was optimized based on an earlier study for its application on LRS (Oliva, 1997). The goal was to simultaneously obtain four polarization states of the FoV/spectra. A simple Wollaston polarizer outputs two polarization states so two Wollaston were used. A special wedge was glued in front of them to tilt and separate the corresponding images onto the CCD (see Fig.\,\ref{fig:DW}). A $80 \times 6$\,mm$^2$ field stop was set on the F/11 FP in photo-polarimetric mode to obtain four $430 \times 33$\,arcsec$^2$ separated FoV strips on the detector. A $0.7 \times 33$\,arcsec$^2$ slit was set in the same focal plane to work in spectro-polarimetric mode. 

The adopted configuration allows us to measure the polarization status of any, even rapidly varying, source in ``one-shot" at the expense of some unavoidable aberration in the PSF shape. We can easily compensate for this aberration for the analysis of point-like object. But the aberration makes the instrument less suitable for extended source polarimetric studies. DW polarimeters also make it easier to correct the time-dependent instrumental polarization that is an intrinsic feature of Nasmith focus polarimeters, as discussed later in Sect\,\ref{sec:mod}.

\begin{figure}
\includegraphics[width=\columnwidth]{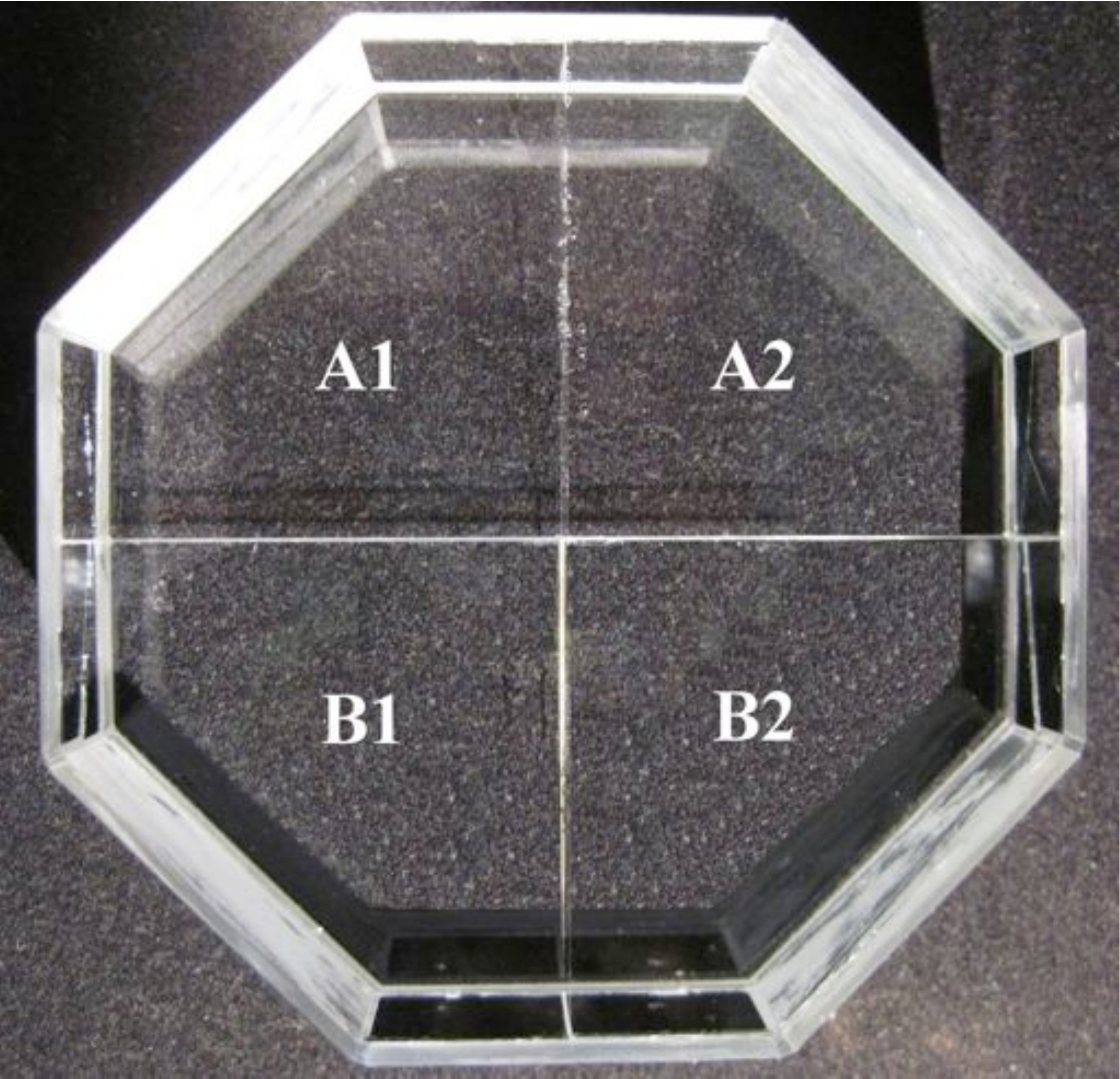}
\caption{The final DW of PAOLO. A1/A2 and B1/B2 crystals pairs, glued to each other, form the two Wollaston with 0-90 and 45-135 polarization states, respectively. The tilting wedges were then glued in front of them to separate the corresponding spot pairs. Then the two blocks are then glued to form the final DW.}
\label{fig:labels}
\end{figure}

The 73\,mm collimated beam exiting from GECO enters the DW. In principle, two $73 \times 35$\,mm Wollaston glued to each other should have been the best solution but it was difficult and expensive to find polarizing crystals of these sizes. Therefore the final choice was to glue two $45 \times 45$\,mm identical Wollastons to form the 0-90 block and other two Wollastons to form the 45-135 one (see Fig.\,\ref{fig:labels}). 

\begin{figure}
\includegraphics[width=\columnwidth]{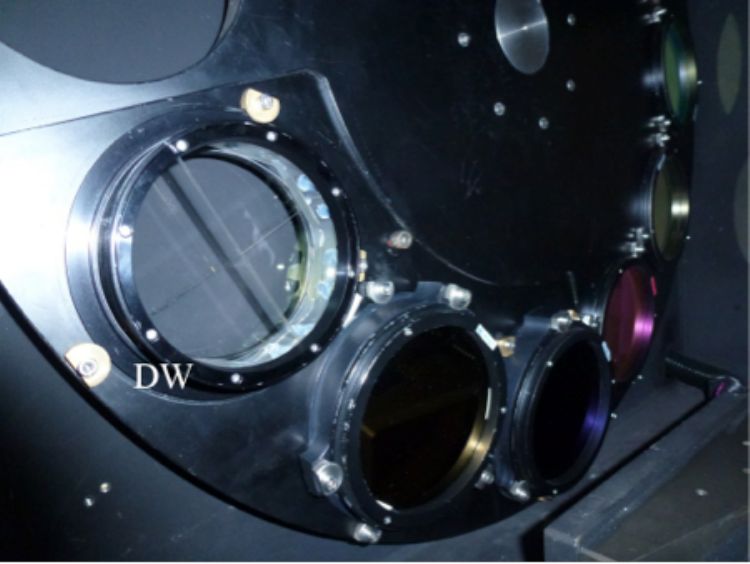}
\caption{The DW mounted on the LRS filter wheel at TNG.}
\label{fig:wheel}
\end{figure}

An hexagonal shape has been obtained to mount the DW in the mechanical interface to the LRS filter wheel (see Fig.\,\ref{fig:wheel}).

A test campaign has been made in optical laboratory. The A1/A2 and B1/B2 crystals have been separately analyzed with a collimated entering beam. The corresponding exiting collimated beam pairs have been focused and analyzed to single out aberration effects. On-sky images show a 0.2\,arcsec (1 pixel) difference in their FWHMs. Throughput tests showed a 95\% transmittance by the DW except in few small bubbled areas where it decreases by less than a 2\%.




%

A setup with a polaroid and a rotating $\lambda$/2 waveplate in front of the DW has allowed us to measure the orientation between the polarization states of its crystal blocks. The beam exiting from $\lambda$/2 waveplate has been collimated by means of a simple optical setup that outputs a $\sim 40$\,mm collimated beam centered on the DW (the intersection point of all the glued faces). The four polarized beams coming out the DW have been focused on the CCD by means of another optical setup. The four spot intensities have been measured for each rotation step of $\lambda$/2 waveplate. There is a 89.89$^\circ$ angle between 0-90 beams. A 90.29$^\circ$ one between 45-135 beams and 45.13$^\circ$ angle between the two couples.

%

\section{PAOLO polarimetric model}
\label{sec:mod}

From the opto-mechanical point of view, PAOLO is a relatively standard instrument. However, it has been designed to work at one of the Nasmyth foci of the TNG.
Polarimeters at a Nasmyth focus are rather uncommon since they require a careful modeling of the optical path in order to remove the important, often dominant, instrumental polarization.

On the other hand, considerable experience in the management of polarimeters located at Nasmyth foci has been gained in recent years, as demonstrated by the publications devoted to the subject: e.g. Giro et al. (2003), Joos et al. (2008), Selbing (2010), Tinberger (2007), Witzel et al. (2011). 

We do not discuss here in detail the various pros- and cons- of placing instruments at a Nasmyth rather than at a Cassegrain focus. Clearly, factors related to mass distribution, control of flexures, etc. play a fundamental role. Polarimetric instruments are indeed typically placed on a Cassegrain focus to avoid the modification of the incoming light beam polarization state by folding mirrors. In addition, for an alt-az telescope such as the TNG, the orientation of folding mirrors with respect to the sky reference changes with time due to field rotation and the reference system of the telescope is of course in turn rotated compared to the sky reference by a time dependent quantity, i.e. the parallactic angle (see, e.g., Fig.\,2 in Giro et al., 2003).

The goal of a polarimetric model is to compensate for the instrumental effect enabling one to know the incoming polarization state from the measured polarization on the detector. Both rotations and folding mirrors effects can be modeled building an appropriate set of Mueller matrices. We remind the readers to the quoted references for details about the mathematical treatment. In principle all the parameters (rotation angles, complex refraction index of the metallic mirror surface, etc.) involved in the model can be accurately measured. However, as it is well known (e.g. Giro et al. 2003), metallic mirrors tend to accumulate dust and other polluting factors with time that can modify their reflective performances and induced polarization. The best solution is to leave the complex refraction index be a free parameter of the model and to observe of a suitable number of standard stars well distributed during an observing night. A best-fit solution can then be readily obtained.

\begin{figure}
\includegraphics[width=\columnwidth]{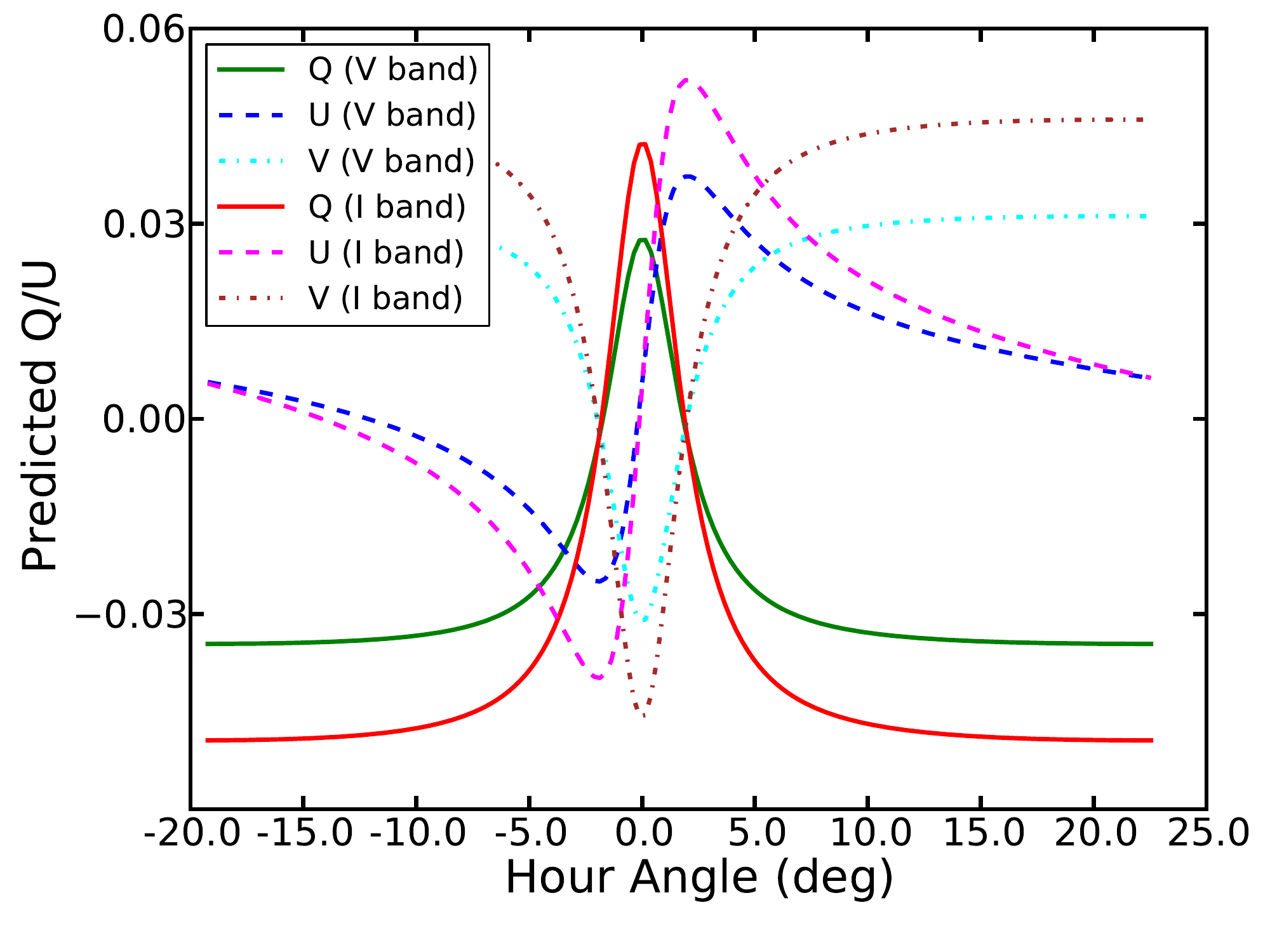}
\caption{Simulation of the instrumental polarization in two filters, $V$ and $I$, introduced by the TNG folding mirror for the observation of a non-polarized standard star BD+303639 on 2012 October 5. For circular polarization a $\lambda$/4 waveplate at 45$^\circ$ was assumed.}
\label{fig:sim}
\end{figure}

The availability of a polarimetric model allows any observer to study in advance the expected instrumental polarization induced by the whole system. We show in Fig.\,\ref{fig:sim} a simulation of the instrumental polarization introduced by the TNG M3 folding mirror for the observation of a non-polarized standard star.
  
Typical values for the instrumental polarization for an unpolarized source observed at three different hour angles are also presented in Table\,\ref{tab:unpol}. The computed circular polarization is obtained assuming to observe a source with 1\% linear polarization and $0^\circ$ position angle, and is totally due to the cross-talk of the $45^\circ$ reflection on the M3 metallic mirror surface.

\begin{table}
\caption{Typical instrumental polarization for an unpolarized source observed at the hour angles: $\sim 0, 15$ and $45^\circ$ in the optical band. The circular polarization is computed for a source with 1\% and $0^\circ$ position angle and no intrinsic circular polarization. The resulting instrumental circular polarization is due to the cross-talks induced by the $45^\circ$ reflection on the M3 metallic mirror.}
\label{tab:unpol}
\begin{tabular}{lccccc}
\hline 
 & $U$ & $B$ & $V$ & $R$ & $I$ \\ 
\hline
Hour angle $\sim 0^\circ$  \\
Polarization (\%) & 2.57 & 2.65 & 2.92 & 3.24 & 4.49 \\
Position angle ($^\circ$) & 6.4 & 6.2 & 5.6 & 4.9 & 3.4 \\
Stokes V (\%) & 0.00 & 0.00 & 0.00 & 0.00 & 0.00 \\
\hline
Hour angle $\sim 15^\circ$ \\
Polarization (\%) & 3.35 & 3.44 & 3.71 & 4.04 & 5.31 \\
Position angle ($^\circ$) & 80.6 & 80.7 & 81.0 & 81.4 & 82.3 \\
Stokes V (\%) & 0.04 & 0.03 & 0.03 & 0.02 & 0.02 \\
\hline
Hour angle $\sim 45^\circ$ \\
Polarization (\%) & 3.09 & 3.17 & 3.44 & 3.77 & 5.03 \\
Position angle ($^\circ$) & -88.1 & -88.0 & -87.5 & -87.0 & -85.7 \\
Stokes V (\%) & -0.09 & -0.08 & -0.07 & -0.06 & -0.05 \\
\hline
\end{tabular}
\end{table}

Once the main parameters of a polarimetric model are determined, for a typical observational run it is sufficient to secure the observation of a small number of polarimetric standard stars in order to update the best-fit values. Our experience at the TNG shows that for a quick-look analysis, models a few weeks old are still perfectly adequate. A set of software tools to manage PAOLO observations, derive a polarimetric model, compensate for instrumental polarization and performing simulations is freely available\footnote{https://pypi.python.org/pypi/SRPAstro.TNG/}. This package, developed for internal use, is offered without any warranty, although we will gladly provide help to any user who requests it. Some more information about how this package works is given in Appendix\,\ref{sec:pao}.

\section{Examples of PAOLO observations}
\label{sec:exa}

\begin{figure}
\includegraphics[width=\columnwidth]{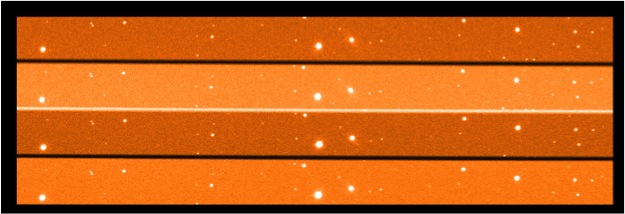}
\caption{The PAOLO field of view, in imagine mode, is formed by four slices providing four position angles for linear or circular polarization measurements. The availability of all the information to properly determine Stokes parameters in one shot is an important feature for instruments affected by rapidly varying instrumental polarization.}
\label{fig:frame}
\end{figure}

The field of view of the PAOLO polarimeter, in imaging mode, is shown in Fig.\,\ref{fig:frame}. To perform imaging polarimetry, one must analyse photometrically the sources of interest in the field of view and then properly match these sources in the four slices to derive instrumental polarization as described, e.g., in Witzel et al. (2011). In case of spectro-polarimetry each spectrum has to be analyzed and wavelength calibrated separately. After that, the procedure is the same, and even for spectro-polarimetry a polarimetric model based on standard star observations has to be derived.

It is possible to roughly estimate PAOLO performances basing on the available DOLORES calibration information\footnote{http://www.tng.iac.es/instruments/lrs/}, simply considering that the collected light is split in four quadrants and taking into due account the expected degree of polarization. However, given the main scientific goal of the project, most of the tests performed so far have been devoted to observations of rapidly variable objects in linear polarimetry rather than sources down to the polarimetric detection limits. Circular polarimetry calibration will be carried out in the near future although in principle the polarization model should work without modifications.

\begin{figure}
\includegraphics[width=\columnwidth]{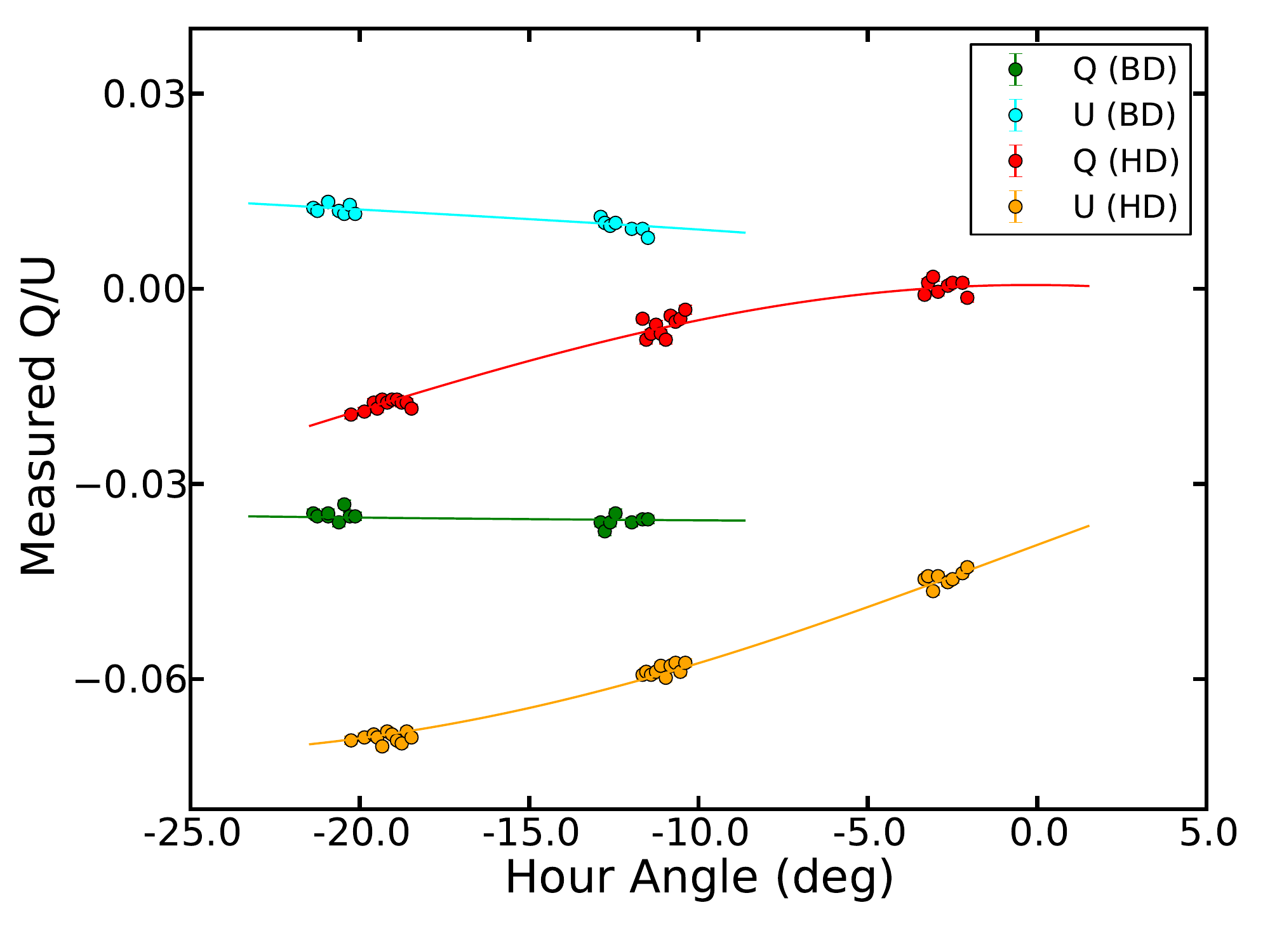} 
\includegraphics[width=\columnwidth]{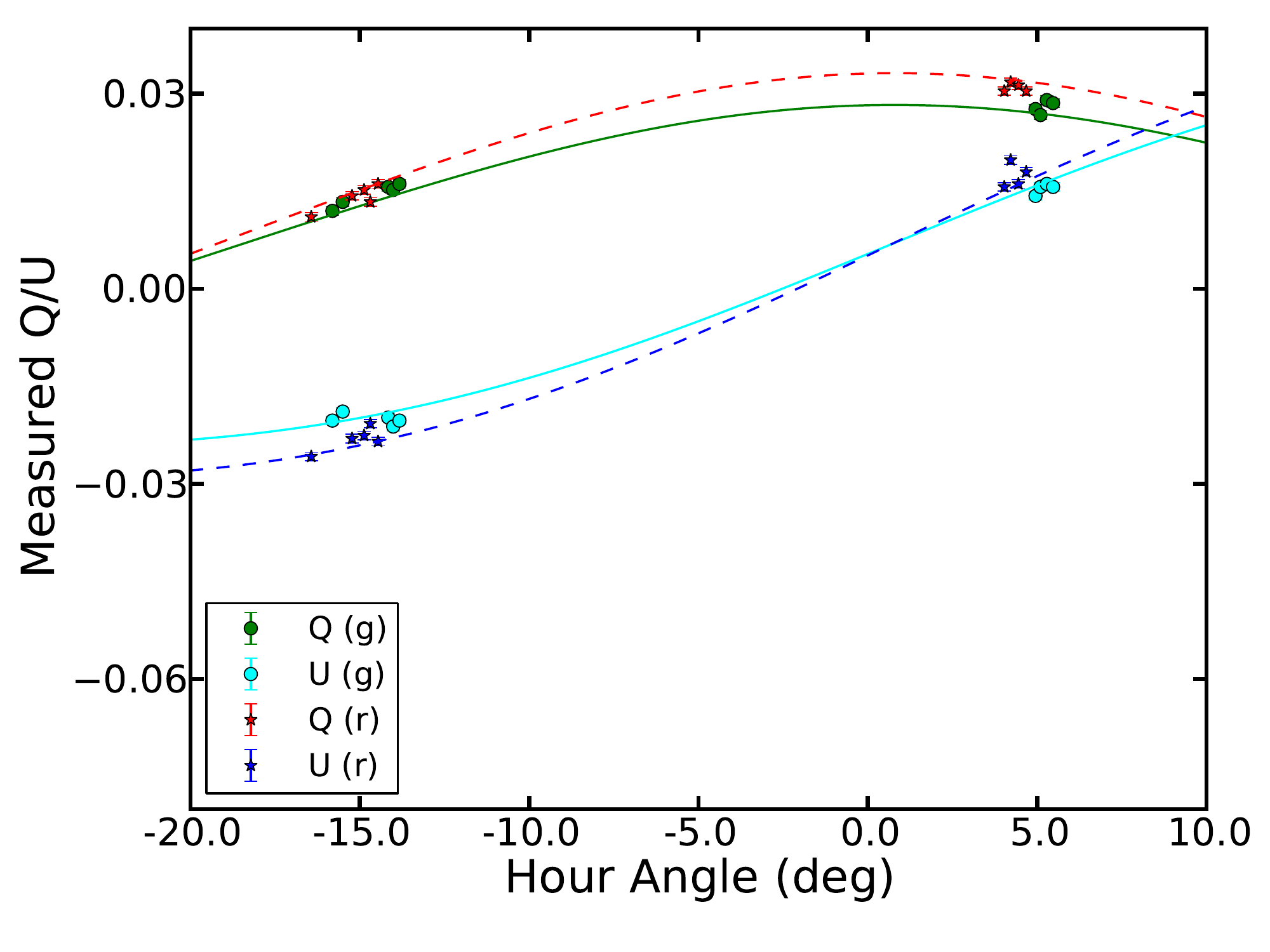} 
\caption{Polarimetric standard star observations (BD+284211 and HD\,204827) carried out on 2012 October 5 (upper panel) and (GD\,319) on 2013 January 27 and 28 (lower panel). We show our best-fit model superposed on the measured instrumental polarizations. The wavelength dependance of the instrumental polarization in this limited range is also well reproduced by the model.}
\label{fig:std1}
\end{figure}

Simultaneously with scientific target observations a few polarized and/or non-polarized standard star observations must be performed. During the commissioning phase we extensively tested the behavior of the instrument for linear polarimetry and some results are shown in Fig.\,\ref{fig:std1}. 

The standard stars observed during these tests are listed in Table\,\ref{tab:std}. Exposure times were typically 1-10\,s depending on the object magnitude and reduction was carried out following standard recipes. Photometric analysis was performed with the DAOPHOT code (Stetson 1987) and $1\sigma$ errors were derived taking into account photon statistics and background determination.

\begin{table}
\caption{Parameters for the standard stars observation discussed in this paper.}
\label{tab:std}
\resizebox{\columnwidth}{!}{
\begin{tabular}{lcccl}
\hline 
Star & Magnitude & Polarization & Position angle & Reference \\ 
       & ($V$ band) & (\%, $V$ band) & ($^\circ$) \\
\hline
BD+284211 & 10.5 & $0.05 \pm 0.03$ & 54.2 & Schmidt et al. (1992) \\
HD\,204827 & 7.9 & $5.39 \pm 0.04$ & 60.0 & Schulz \& Lenzen (1983) \\
GD\,319 & 12.3 & $0.09 \pm 0.09$ & 140.2 & Schmidt et al. (1992) \\
\hline
\end{tabular}}
\end{table}

The errors associated with the catalogue standard star polarizations are of the order of a few $\times 0.01$\%, while the photometric errors for the bright (with a 3.6\, telescope) observed objects for differential measurements are low enough to give Stokes parameters measurements with comparable uncertainties.
The scatter around the best-fit curves reported in Fig.\,\ref{fig:std1} is typically an order of magnitude larger and therefore, from the rms error evaluated comparing the observations to model predictions, we can infer that the removal of instrumental polarization can be performed at the 0.2\% level, although relative measurements can be more accurate (see also Wiltzer et al. 2011). The model best fit components of the complex refractive index of the M3 mirror were typically about 90\% the values tabulated for pure Aluminium\footnote{http://www.filmetrics.com/refractive-index-database/Al/Aluminium}.

\begin{figure}
\includegraphics[width=\columnwidth]{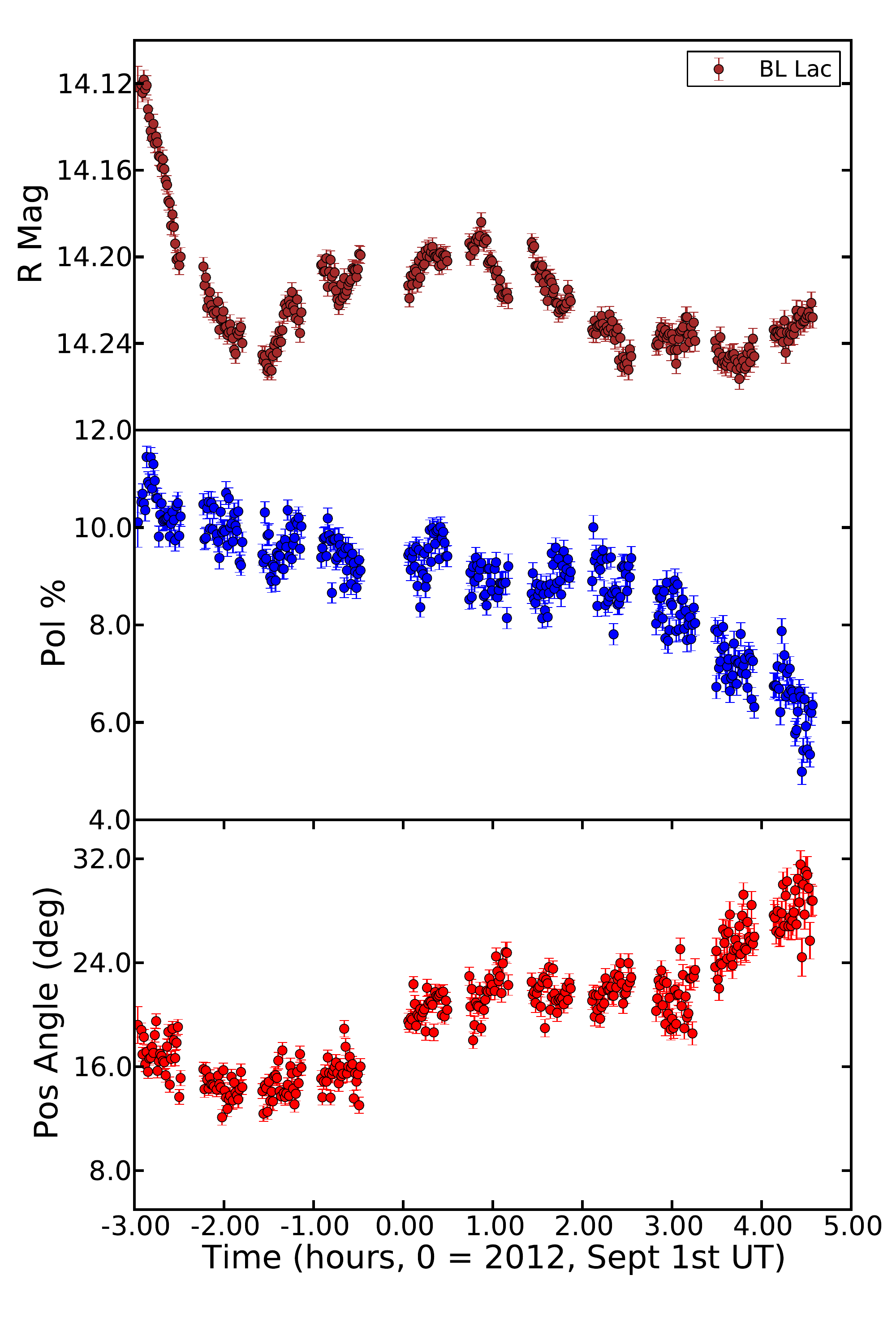}
\caption{Linear polarimetric and photometric monitoring of BL\,Lac carried out on 2012 September 1.}
\label{fig:blc}
\end{figure}

An example of a real scientific observation, although performed as a system test, is shown in Fig.\,\ref{fig:blc}. On 2012 September 1 we carried out a continuous linear polarimetric and photometric monitoring of BL\,Lac. BL\,Lac is one of the most intensively studied AGNs even in polarimetry, and long time-series covering a few decades are available (e.g. Hagen-Thorn et al. 2002). We tried to study its linear polarization characteristics on short time-scales, typical integration times were of 20\,s. The polarization degree during the slightly less than eight hour monitoring decreased from about 11\% to 6\%, while the position angle on the contrary increased from about 15\,$^{\circ}$ to less than 30\,$^{\circ}$. Some relevant short-term activity is indeed present in the data (see an analogous study in Andruchow, Romero \& Cellone, 2005 for other AGNs) and a detailed scientific analysis will be presented elsewhere. 





\section{Conclusions}
\label{sec:conc}

In this paper we have presented a new polarimetric facility, PAOLO, available at the INAF / TNG. TNG is an alt-az telescope and PAOLO is located at one of its Nasmyth foci. This positioning introduces time- and pointing position-dependent polarization. We described the methodology followed to model the instrumental polarization and remove it from scientific data. A set of software tools to analyze PAOLO polarimetric data have also been developed.

A few examples of possible applications have also briefly described. PAOLO is now offered to the general community as part of the instrument set of the INAF /TNG.

\acknowledgements
This work was made possible by a grant from the INAF TecnoPRIN 2009. We thanks G. Pareschi for his continuous encouragement and for having created the acronym of the project. We also thanks L. Foschini and A. Treves for useful discussions and J. Schwarz for for editorial assistance. We thank the anonymous referee as well for her/his very useful comments.

\newpage

\appendix

\section{The PAOLO software tools}
\label{sec:pao}

In order to aid the calibration and commissioning activities for PAOLO we have developed a set of command-line tools for managing the main phases of a typical polarimetric analysis session. These tools are thought to match the PAOLO instrumental features, nevertheless some of them may be of general use or interest. The package is entirely written in the {\tt python} language\footnote{http://www.python.org} and is based on the standard numerical and scientific libraries {\tt Numpy}\footnote{http://www.numpy.org} and {\tt Scipy}\footnote{http://www.scipy.org}. Here we briefly describe how the package works. An user's manual, mainly devoted to installation and with some recipe for analysis is also available\footnote{http://pythonhosted.org/SRPAstro.TNG/}.

\begin{enumerate}
\item PAOLO can work both in imaging and spectro-polarimetric modes. Photometry can be derived by means of any suitable tool, and it is only required that as output a file with as many entries as the number of objects analyzed be produced. Each entry must contain at least an Id, pixel position on the detector and instrumental magnitudes. The {\tt SRPTNGPAOLOSourceMatch} command takes care of matching the different sources in the four quadrants constituting the PAOLO field of view in imaging (see Fig.\,\ref{fig:frame}). In case of spectroscopy, again spectral extraction and wavelength calibration can be carried out with any suitable tool. The {\tt SRPTNGPAOLOSpectrumMatch} imports the extracted spectra and rebins them if required in order to have the same number of pixels. Both commands also compute the total magnitude for each detected source and the sum of fluxes for spectroscopy. A FITS\footnote{http://fits.gsfc.nasa.gov} table managed with the {\tt ATpy} tool\footnote{http://atpy.readthedocs.org/en/latest/} containing all the imported data is generated as output.

\item Once data are properly imported instrumental polarization Stokes parameters are computed with the command {\tt SRPTNGPAOLOInstrStokes} following standard recipes (e.g. Giro et al. 2003). Errors are correctly propagated. In case one is managing observations of polarimetric standard stars catalogued polarization Stokes parameters can be provided. In addition parameters from the original observation frames such as pointing direction, epoch, exposure times, etc. are obtained.

\item The most important step is the derivation of the best-fit parameters for the instrument polarimetric model (see Sect\,\ref{sec:mod}). The command {\tt SRPTNGPAOLOParamFit} will perform this fit using one or more parameters as chosen by the user. At present the available parameters are the offset in the detector position angle, real and complex refraction indices of the metallic mirror surface, and Stokes parameter instrumental polarization. The best-fit is computed by $\chi^2$ minimization by using the downhill (Nelder-Mead) simplex algorithm as coded in the {\tt Scipy} library, v.\,0.12.0. This is not the most efficient algorithm, yet it is simple and easy to manage.

\item Once best-fit parameters for the polarimetric model are obtained, one can remove the time-dependent instrumental polarization with the command {\tt SRPTNGPAOLOCalStokes}. It generates a new table with instrumental and corrected Stokes parameters, together with information about the pointing direction, epoch, etc. for each epoch.

\item There is also the possibility to generate simulated sets of instrumental polarization in order to compare with standard star observations, plan future observations, etc. The command is {\tt SRPTNGPAOLOStokesSim}.
\end{enumerate}

The package is part or a larger set of tools we developed for generic astronomical data analysis\footnote{https://pypi.python.org/pypi/SRPAstro}. Again these tools are freely available, although they have been developed without following a specific strategy, i.e. to solve detailed and often unrelated problems in dealing with astronomical datasets.

\end{document}